\documentclass{article}
\usepackage{epsfig}
\usepackage{amsmath}
\usepackage{amssymb}

\newcommand{\longpage}{\enlargethispage{\baselineskip}}
\begin{document}

\title{CP violation in a light Higgs boson decay from {\boldmath$\tau$\unboldmath}-spin correlations at a linear collider}
\author{ Andr\'e Roug\'e
\thanks{Andre.Rouge@in2p3.fr}\\
Laboratoire Leprince-Ringuet, Ecole Polytechnique-IN2P3/CNRS,\\
 F-91128 Palaiseau Cedex}
\date{31 May 2005}
\maketitle
\begin{abstract}
We present a new method to measure the transverse spin correlation in the
\mbox{$H\to\tau^+\tau^-$} decay. The method has been devised to be insensitive to the beamstrahlung
which affects  the definition of the beam energy at a linear collider.
In the case of two $\tau^\pm\to\pi^\pm\bar{\nu}_\tau(\nu_\tau)$ decays, using the anticipated  detector 
performance of the TESLA project, we get a promising estimation of the error 
expected 
on the measurement of a CP violating phase.
\end{abstract}
\section{Introduction}
\label{sec:intro}

The possibility to determine the CP properties of a light Higgs boson through 
the spin correlations in its $H\to\tau^+\tau^-$ decay has  been 
often considered~[1--9]. 
The principle is simple. 
Let  $\pm$ denote the projection of the spins of the $\tau$'s in their respective 
rest frames on a $z$-axis oriented in the  direction of the $\tau^-$ line of flight for the $\tau^+$
and opposite to the $\tau^+$ line of flight for the $\tau^-$ . The $\tau^+\tau^-$ spin state
\begin{equation}\label{eq:etat}
\frac{1}{\sqrt{2}}[|+-\rangle +e^{i\xi}|-+\rangle]
\end{equation}
is a CP$=+1$ state for $\xi=0$, a CP$=-1$ state for $\xi=\pi$, and a mixed CP state otherwise.
Such a state is produced by the decay of a CP=+1 Higgs with a coupling~\cite{ssd,wc}
\begin{equation}
g\bar\tau (\cos\psi+i\sin\psi\gamma_5)\tau H~.
\end{equation}
In  this case, neglecting $\mathcal{O}(m_\tau^2/m_H^2)$, the phase is $\xi=2\psi$.

The spin correlations for the state (\ref{eq:etat}) are:
\begin{equation}
C_{zz}=-1,\hspace{.5em}C_{xx}=C_{yy}=\cos\xi=\cos2\psi,
\hspace{.5em}C_{xy}=-C_{yx}=\sin\xi=\sin2\psi~.
\end{equation}
The way to measure $\psi$ ($\xi$)  is transparent in the case of two
$\tau^\pm\to\pi^\pm\bar\nu_\tau(\nu_\tau)$ decays.
Using (\ref{eq:etat}) and the $\tau$ decay amplitudes, one gets the correlated decay distribution
\begin{equation}\label{eq:w3d}
W_3(\cos\theta^+,\cos\theta^-,\Delta\varphi)=\frac{1}{8\pi}\left[1+\cos\theta^+\cos\theta^-
-\sin\theta^+\sin\theta^-\cos(\Delta\varphi-2\psi)\right]~,
\end{equation}
where $\theta^\pm$ is the polar angle in the $\tau^\pm$ rest frame between the $\pi^\pm$ 
direction ($\hat\pi$)  and the above defined $z$-axis. The relative azimuthal angle 
$\Delta\varphi=\varphi^+-\varphi^-$ is the angle between the two 
planes defined by the $\tau^-$ direction and the $\pi^+$ ($\pi^-$) direction respectively  in the Higgs rest frame.
The distribution of the azimuthal angle is obtained by integrating out the polar angles:
\begin{equation}\label{eq:w1d}
W_1(\Delta\varphi)=\frac{1}{2\pi}\left[1-\frac{\pi^2}{16}\cos(\Delta\varphi-2\psi)\right]~.
\end{equation}
The merits of the two distributions for the measurement of $\psi$ can be quantified by 
their sensitivities $S_\psi=1/\sigma_\psi\sqrt{N}$, where $\sigma_\psi$ is the error 
on $\psi$ expected from a maximum likelihood fit of the distribution for a sample of 
$N$ events.
The sensitivities measure the information  per event on $\psi$ contained in the 
distributions~\cite{omega}; their computation is straightforward when an analytical 
or numerical expression of the distributions is known.
They are $S_\psi^3=1.15$
for the distribution~(\ref{eq:w3d}) and $S^1_\psi=0.92$ for~(\ref{eq:w1d}).
The superiority of  (\ref{eq:w3d}) is not very large and it decreases when experimental effects 
are introduced. For that reason and for the sake of simplicity, we will only consider the distribution~(\ref{eq:w1d}) in the following.

The best place to study a light Higgs at a linear $e^+e^-$ collider is the 
Higgsstrahlung process $e^+e^-\to ZH$. We assume that the mass of the Higgs 
is well measured by the analysis of its dominant decay modes. The four-momentum
 of the Higgs is  determined by the measurement of the Z and therefore the 
four-momenta of the $\tau^+$ and the $\tau^-$ can, in principle, be reconstructed in the case of two hadronic decays (Sec.~\ref{sec:hgfr}). 
It is known~\cite{omega} that under such circumstances the three main hadronic decay modes,  $\tau^\pm\to\pi^\pm\bar{\nu}_\tau(\nu_\tau)$,
$\tau^\pm\to\rho^\pm\bar\nu_\tau(\nu_\tau)$, and $\tau^\pm\to a_1^\pm\bar{\nu}_\tau(\nu_\tau)$  have the same analysis power for the 
measurement of spin effects. This is due to the fact~[11,13]
that, in the 
$\tau$ rest frame, all the information on the  spin is embodied in the distribution of 
a unit vector $\hat a$, the polarization analyser, which is equal to $\hat\pi$
(pion direction) in the case of  $\tau^\pm\to\pi^\pm\bar{\nu}_\tau(\nu_\tau)$. This vector can be 
computed~\cite{kw,analv} from the measured momenta in the case of the above decay modes. 
For example, if $\tau^\pm\to\rho^\pm\bar\nu_\tau(\nu_\tau)$ and $\rho^\pm\to\pi^\pm\pi^0$, 
then
\begin{equation}
a^i=\mathcal{N}\left(2(q\cdot p_\nu)q^i-(q\cdot q)p_\nu^i\right)~,
\end{equation}
where $\mathcal{N}$ is a normalization factor, $p_\nu=p_{\tau^\pm}-p_{\rho^\pm}$ is the four-momentum 
of the neutrino, and $q=p_{\pi^\pm}-p_{\pi^0}$ is the difference of the four-momenta of the
two pions.
The distribution (\ref{eq:w3d}), where $\hat \pi$ is replaced by $\hat a$ in the 
definition of the angles, contains all the available information on the 
spin correlation. The three decay modes (i.e., (55\,\%)$^2$ of the $\tau^+\tau^-$ pairs) 
can therefore be used in the same way for the 
measurement of the CP violating parameter $\psi$~\cite{ssd}.

Unfortunately, it has been shown by Monte Carlo studies~\cite{ww} that the 
standard reconstruction (Sec.~\ref{sec:hgfr}) of the $\tau$ four-momenta is critically impaired
by the effects of beamstrahlung. A new estimator has been 
proposed~[7--9], which is less sensitive to the quality of the 
$\tau$ reconstruction but requires that both $\tau^+$ and $\tau^-$ decay by the process
 $\tau^\pm\to\rho^\pm\bar\nu_\tau(\nu_\tau)$. However, the ideal sensitivity of the new method,
when all the four-momenta are exactly known, is $0.48$ to be compared with $0.92$ for the standard 
method. Besides it takes advantage of 
$(25\%)^2$ only of the $\tau^+\tau^-$ pairs. There is therefore a possible improvement of the error 
on $\psi$ by a factor of at least four, if a method of reconstruction of the $\tau$'s insensitive to 
beamstrahlung is found.   

The aim of the present paper is to devise such a method. To check the sensitivity to 
beamstrahlung, we use a simple Monte Carlo procedure.  First the energies of the $e^\pm$ after 
beamstrahlung are generated using the program {\sf circe}~\cite{circe} with the parameters of 
the TESLA project and the cross section for Higgsstrahlung in the case of a CP=+1 
Higgs~\cite{ssc}; next the production and decay angles of the Z are generated according to their 
correlated distribution under the same hypothesis~\cite{ssc}. 
The Higgs decay angles are generated according to an isotropic distribution. For the correlated decay of the 
$\tau$'s we consider three cases: two decays into $\pi\nu$ ($\pi\pi$), a decay into $\pi\nu$ and 
a decay into $\rho\nu$ ($\rho\pi$), and two decays into $\rho\nu$ ($\rho\rho$). For each case, 
the azimuthal angles and the cosines of the polar angles in the decays of $\tau$'s and $\rho$'s
are generated uniformly. From Eq.~\ref{eq:etat} and the decay amplitudes for $\tau^\pm\to\pi^\pm\bar{\nu}_\tau(\nu_\tau)$ and
 $\tau^\pm\to\rho^\pm\bar\nu_\tau(\nu_\tau)\to\pi^\pm\pi^0\bar\nu_\tau(\nu_\tau)$, the correlated decay probabilities are computed for diverse values 
of $\psi$ and the events are accepted or rejected accordingly. The four-momenta of all 
the particles are then computed and, finally, the decay lengths of the $\tau$'s are generated.

The Higgs mass is assumed to be 120~GeV and three values of the total energy are 
considered: $\sqrt{s}=230~\mathrm{GeV}$, $\sqrt{s}=350~\mathrm{GeV}$ 
and $\sqrt{s}=500~\mathrm{GeV}$. The natural energy for a detailed study of the Higgs is 
the energy of the largest cross section (i.e. near  $230~\mathrm{GeV}$). The consideration of higher 
energies allows  estimating the robustness of the method and possibly the effect of an 
underestimation of radiative effects.
\section{Reconstruction of the {\boldmath$\tau$\unboldmath}'s in the Higgs rest frame} \label{sec:hgfr}
\begin{figure}[t]
  \begin{center}
    \epsfig{file=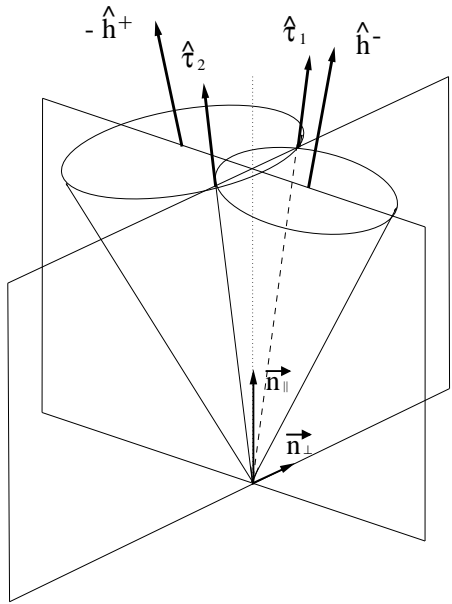,width=6.5cm}
    \caption[.]{Reconstruction of the  $\tau$ direction in 
the Higgs rest frame.
    \label{fig:taucone}}\normalsize

  \end{center}
\end{figure}

The reconstruction of the $\tau$'s in the $\tau^+\tau^-$ rest frame for hadronic decays
$\tau^\pm\to h^\pm\bar\nu_\tau(\nu_\tau)$ has been known for a long time~\cite{tsh} and 
was used at LEP to improve the measurement of the $\tau$ polarization~\cite{polal}. Its principle
is sketched in Fig.~\ref{fig:taucone}. 
Both $\tau$'s have the same energy $E_\tau^\pm=m_H/2$ and the energies and momenta of 
the hadrons are measured. The angle $\delta^\pm$ between the direction of a $\tau^\pm$ displayed by the unit 
vector $\hat\tau^\pm$ and the direction of the hadron $\hat h^\pm$ is therefore fixed:
\begin{equation}\label{eq:cosdelta}
\cos\delta^\pm=\frac{2E_{\tau^\pm}E_{h^\pm}-m_\tau^2-m_{h^\pm}^2}{2p_{\tau^\pm}p_{h^\pm}}~.
\end{equation}
The direction of the $\tau^-$, $\hat\tau\equiv\hat\tau^-$ is on the intersection between the cone around
$\hat h^-$ with angle $\delta^-$ and the cone around $-\hat h^+$ with angle $\delta^+$.

In general there are two solutions 
$\hat\tau^{1/2}=\vec{n}_{\scriptscriptstyle\parallel}\pm\vec{n}_\perp$, 
where the vectors $\vec{n}_{\scriptscriptstyle\parallel}$ and $\vec{n}_\perp $, 
shown in Fig.~\ref{fig:taucone},  are given by \begin{eqnarray}\label{eq:nparperdef}
\vec{n}_{\scriptscriptstyle\parallel}&=&-\frac{\cos\delta^+-\cos\delta\cos\delta^-}{\sin^2\delta}\,\hat{h}^+
+\frac{\cos\delta^--\cos\delta\cos\delta^+}{\sin^2\delta}\,\hat{h}^-~,\\[0.5ex]\nonumber
\vec{n}_\perp&=&\sqrt{\frac{1-\vec{n}_{\scriptscriptstyle\parallel}\,\!\!^2}{\sin^2\delta}}\,\hat{h}^+\wedge\hat{h}^-~,
\hspace{1em}\text{with}\hspace{1em}\cos\delta=-\hat{h}^+\cdot\hat{h}^-~.
\end{eqnarray}
The ambiguity can be resolved using the information from a vertex detector but, because the main 
vertex is known from the Z decay, the use of the detector is rather different and simpler than at
 LEP~\cite{polal,kuhn93}. One needs only to make a chi-square test on the distances  in the laboratory
between the reconstructed $\tau$ lines of flight and the trajectories of the charged pion's. 

The situation is degraded in the presence of beamstrahlung and/or other experimental effects 
because the intersection of the two cones is no more granted. As a result the acceptance is 
reduced and the distribution
 of $\Delta\varphi$ strongly deformed. 
The reconstruction is especially awkward in the case of two  $\tau^\pm\to\pi^\pm\bar{\nu}_\tau(\nu_\tau)$ decays, because the polarization 
analyser is then $\hat{a}^\pm=\hat{\pi}^\pm=\hat{h}^\pm$ and thus $\Delta\varphi$ is  the angle 
between the vectors $\hat{\tau^-}\wedge\hat{h}^-$ and $\hat{\tau^-}\wedge\hat{h}^+$. 
From Eq.~\ref{eq:nparperdef}, one gets
\begin{equation}
1-\cos^2\Delta\varphi=(1-\vec{n}_{\scriptscriptstyle\parallel}\,\!\!^2)\frac{\sin^2\delta}{\sin^2\delta^+\sin^2\delta^-}~,
\end{equation}
which shows that the two cones are tangent when $|\cos\Delta\varphi|=1$, feature that can also be observed in Fig.~\ref{fig:taucone}.
As a consequence, even for $\sqrt{s}=230\,\mathrm{GeV}$, where the closeness to the threshold reduces the effect of 
beamstrahlung, the acceptance becomes very small when $\cos\Delta\varphi$ is near $\pm 1$ and the
 distribution~(\ref{eq:w1d}) can hardly be used to test the CP properties of the Higgs.
\section{Reconstruction of the {\boldmath$\tau$\unboldmath}'s in the laboratory}
\label{sec:lab}
Owing to the observation of the Z decay products, which allows the reconstruction of  the main vertex, 
one may envisage to perform the reconstruction of the $\tau$'s in the laboratory frame.

\begin{figure}[htp]
\begin{center}
    \epsfig{file=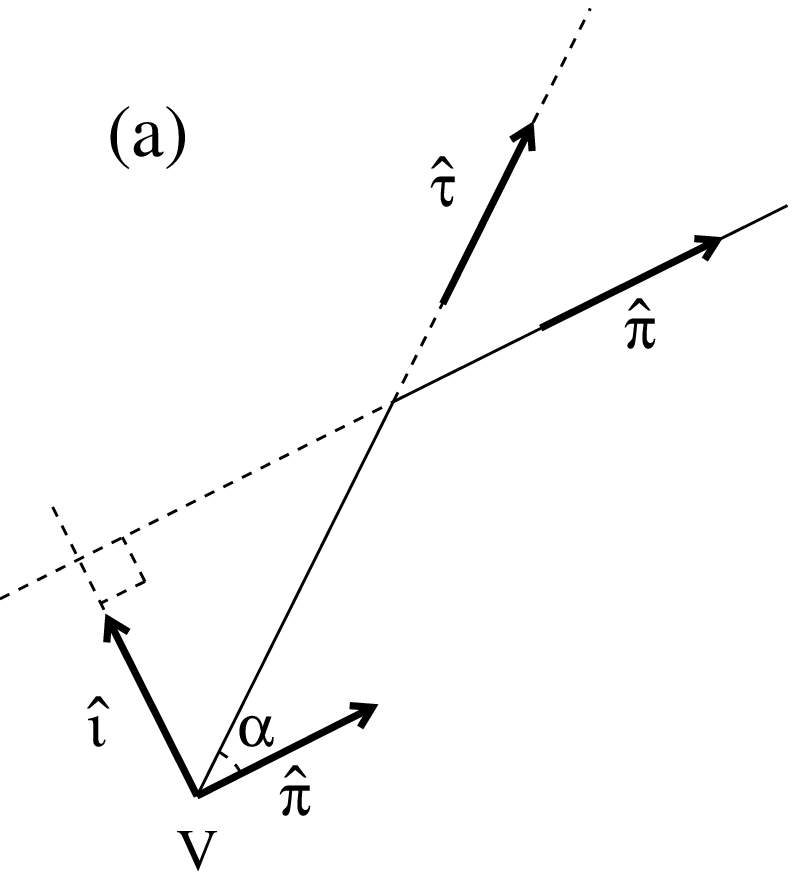,width=4.5cm}\hspace{2cm}
    \epsfig{file=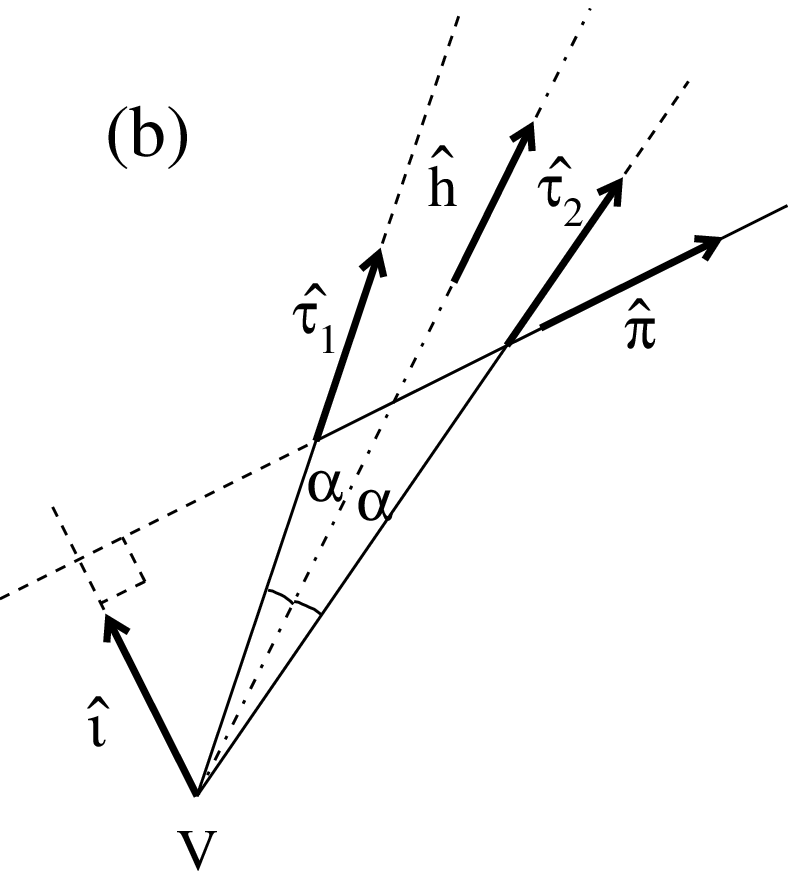,width=4.5cm}
    \caption[.]{Reconstruction of the $\tau$ direction in the laboratory frame:
(a) for a $\tau^\pm\to\pi^\pm\bar\nu_\tau(\nu_\tau)$ decay mode, (b) for a 
$\tau^\pm\to\pi^\pm \pi^0\bar\nu_\tau(\nu_\tau)$ decay mode. \label{fig:taulab}}
\end{center}
\end{figure}

Let us assume that the  $\tau^\pm$ energies in the laboratory $E_{\tau^\pm}^L$ are known. The angles
$\alpha^\pm$ between the directions of the $\tau$'s and the hadrons are given by the 
relation~(\ref{eq:cosdelta}), which reads here   
\begin{equation}\label{eq:cosalpha}
\cos\alpha^\pm=\frac{ 2E_{\tau^\pm}^LE_{h^\pm}^L-m_\tau^2-m_{h^\pm}^2}{2p_{\tau^\pm}^Lp_{h^\pm}^L}~.
\end{equation}
The reconstruction of the $\tau^\pm$ direction is then very simple in the case of the 
 $\tau^\pm\to\pi^\pm\bar\nu_\tau(\nu_\tau)$ decay mode (Fig.~\ref{fig:taulab}-a).

 Denoting by $\hat{\iota}$ the unit vector of the 
perpendicular from   the vertex to the pion trajectory and by $\hat\pi$, 
the unit vector of the pion momentum, the  vector $\hat\tau$ is given by
\begin{equation}
\hat\tau=\cos\alpha\,\hat\pi+\sin\alpha\,\hat{\iota}~.
\end{equation}
In the case of a $\tau^\pm\to\pi^\pm \pi^0\bar\nu_\tau(\nu_\tau)$ decay mode, $\hat\tau$ is 
on the intersection of the plane defined by the vertex
and the charged $\pi$ trajectory with
the cone around $\hat h$ with angle $\alpha$  (Fig.~\ref{fig:taulab}-b). 
There are in general two solutions. 
The problem of the reconstruction of the two $\tau$'s is therefore a problem with two unknowns: 
$E_{\tau^+}^L$ and $E_{\tau^-}^L$, but even in the presence of beamstrahlung we still have three 
constraints: the conservation of the components of the momentum orthogonal to the beams and the 
equality of the $\tau^+\tau^-$ effective mass with the Higgs mass. They are sufficient to determine 
the two energies and resolve the ambiguities if needed.
\section{A simplified algorithm}\label{sec:sa}
\begin{table}[b]
  \begin{center}
    \caption[.]{ The sensitivities to $\psi$ of the reconstructed distributions when only 
beamstrahlung is taken into account. The effect of the small loss of acceptance in the
$\pi\pi$ channel is included. \label{tbl:sensbsonly}
         }

     \begin{tabular}{lccc} \\\hline
    \raisebox{-1.5ex}[-1.5ex]{$\sqrt{s}~(\mathrm{GeV})$} &   \multicolumn{3}{c}{Sensitivity ($S_\psi$)}\\
	&$\pi\pi$&$\pi\rho$&$\rho\rho$  \\ \hline
230&0.92&0.88&0.83\\350&0.91&0.73&0.66\\
500&0.88&0.64&0.55\\
\hline
    \end{tabular}
\end{center}
\end{table}
Implementing the last method by a fit would require a good knowledge of the errors and 
their correlations. This is not possible with our simple simulation. For that reason,
we will use a new procedure, which combines elements of the approaches followed in the two 
previous sections and gives good results, without the intricacies of a fit.  

Taking $\vec{p}_H=-\vec{p}_Z$, we start the reconstruction in the Higgs rest frame but use 
the approximation 
$\hat\tau=\vec{n}_{\scriptscriptstyle\parallel}/|\vec{n}_{\scriptscriptstyle\parallel}|$. 
This is always possible if $\cos\delta^\pm$ is replaced by $1$ when it is found greater than 1.
\begin{figure}[t]
\begin{center}
    \epsfig{file=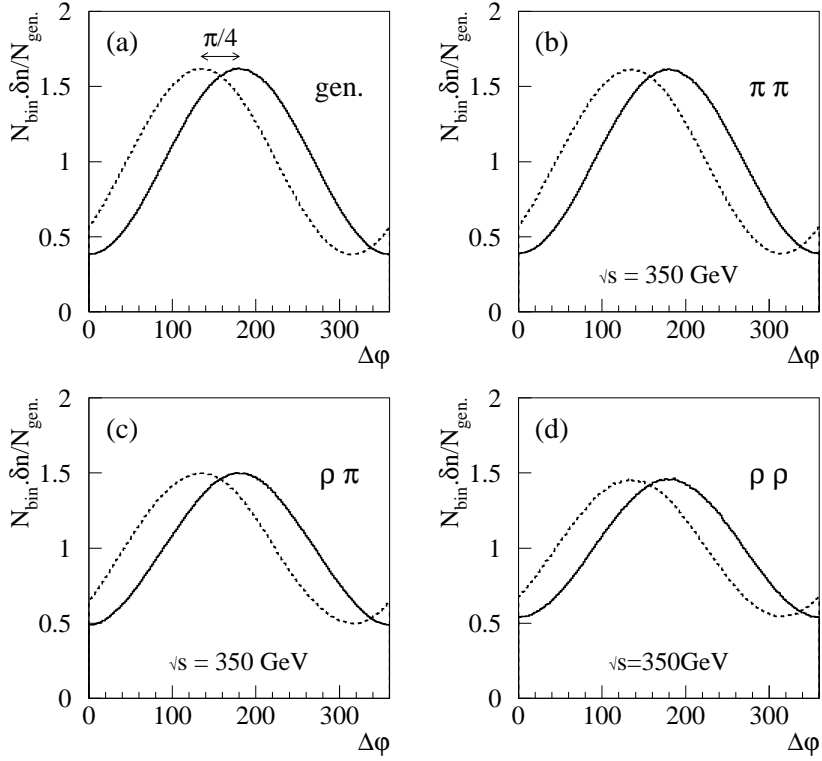,width=12cm}
    \caption[.]{
The distributions of $\Delta\varphi$: (a) at the generation level, (b), (c), and (d), 
reconstructed by the method of Sec.~\ref{sec:sa} for the three channels $\pi\pi$,
$\rho\pi$, and $\rho\rho$ at an energy of 350 GeV. Beamstrahlung effects only are taken into account.
The histograms are normalized to the number of generated events and multiplied by the number of bins. With this normalization, the distribution (a) is $2\pi W_1(\Delta\varphi)$.
The full lines correspond to $\psi=0$, the dotted lines to $\psi=\pi/8$.\label{fig:bsonly}}
\end{center}
\end{figure}

As the $\tau$ energies in the Higgs rest frame are known: $E_{\tau^\pm}=m_H/2$, we can compute 
$E_{\tau^\pm}^L$ and perform the $\tau$ reconstruction in the laboratory. For that, we replace
$\cos\alpha^\pm$  by $1$ when it is found greater than 1 and, in the case of a $\tau^\pm\to\rho^\pm\bar\nu_\tau(\nu_\tau)$ decay,
 use the projection of $\hat h$ on the (V,$\hat\pi$,$\hat\iota$) plane to define $\hat\tau$ when the cone 
and the plane do not intersect. In the $\pi\pi$ channel, an event is rejected if both
$\cos\alpha^+$ and $\cos\alpha^-$ are greater than 1. The ambiguities are resolved by choosing 
the solution with the smallest missing $|\vec{p}_\perp|$.
We can now redefine the Higgs frame as the rest frame of the $\tau^+\tau^-$ pair and compute
$\Delta\varphi$.\longpage

The reconstructed distributions for the three channels at an energy of
$\sqrt{s}=350\,\mathrm{GeV}$ are shown in Fig.~\ref{fig:bsonly} and their 
sensitivities to $\psi$ at the three considered energies are given in Table~\ref{tbl:sensbsonly}.

Both the curves in Fig.~\ref{fig:bsonly} and the numbers in Table~\ref{tbl:sensbsonly} include  the
effect of the small loss of acceptance due to the rejection of events in the $\pi\pi$ channel.

For the $\pi\pi$ channel, the sensitivity is nearly the ideal one (0.92) up to 500\,GeV.
For the $\rho\pi$ and $\rho\rho$ channels, the sensitivities are slightly reduced at 230\,GeV and 
decrease more rapidly with the energy than for the $\pi\pi$ channel. Two effects contribute to 
that. The first is the closing of the $\tau$ decay angle when the hadron mass increases. The second is 
the imperfect resolution of the ambiguities. The second point can be improved because the 
criterion used for the choice of the solution is not optimal. 
Other constraints can be used, like the $\tau^+\tau^-$ effective mass  and 
the positivity of the decay length. It should also be noted that the conservation 
of $\vec{p}_\perp$ is an important but not vital point in the method. For example, 
smearing the $\vec{p}_\perp$ of the Z with $\sigma(p_x)=\sigma(p_y)=1\,\mathrm{GeV}$ yields
sensitivities $S_{\rho\pi}=0.66$ and $S_{\rho\rho}=0.52$ at $\sqrt{s}=350\mathrm{GeV}$.

Finally, a few remarks are in order about the $\tau^\pm\to a_1^\pm\bar{\nu}_\tau(\nu_\tau)$ decay channel, which is not 
included in our unsophisticated simulation, for the reason that its description is more complex 
than a simple angular distribution.
The $a_1^\pm\to\pi^\pm\pi^0\pi^0$ decay mode is reconstructed by the same method that the 
$\rho^\pm$, but the sensitivity will probably be worsened by the closing of the $\tau$ 
decay angle. For the  $a_1^\pm\to\pi^\pm\pi^+\pi^-$ decay mode, the determination of $\hat\tau$
in the laboratory is in principle possible from vertexing information only, consequently  a good 
sensitivity can certainly be achieved by using this determination of $\hat\tau$
or by an adaptation of the method used for the $\rho^\pm$.
\section{A semi-realistic simulation of the {\boldmath$\pi\pi$\unboldmath} channel}
\label{sec:sms} 
To get a realistic estimation of the sensitivities, it is necessary to take into account 
the performance  of the detector. We use for that the parameters of the TESLA 
project~\cite{teslaa,teslab}. Since the precision of the $\pi^0$ measurement depends not only on 
the accuracy of the detector but also  on the quality of the reconstruction algorithm, 
we consider here the $\pi\pi$ channel only.
\begin{table}[b]
  \begin{center}
    \caption[.]{The sensitivities to $\psi$ of the reconstructed distributions for the $\pi\pi$ channel, when all the experimental effects are taken into account.
             \label{tbl:pipismr}
}
    \vspace{0.3cm}
    \begin{tabular}{lcc} \hline
    \raisebox{-1.5ex}[-1.5ex]{$\sqrt{s}~(\mathrm{GeV})$} &   \multicolumn{2}{c}{Sensitivity ($S_\psi$)}\\
	&$Z\to\mu^+\mu^-$&$Z\to q\bar q$  \\ \hline
230&0.69&0.71\\350&0.60&0.61\\
500&0.58&0.58\\
\hline
    \end{tabular}
\end{center}
\end{table}

For the charged tracks an independent Gaussian smearing is performed on the five parameters: 
$\theta$, $\phi$, $1/p_\perp$ and  the two components of the impact parameter. We  
use for the widths of the Gaussians the following values~\cite{teslab}: $$\sigma(\theta)=\sigma(\phi)=0.1~\mathrm{mrad},\hspace{2em}  
\sigma(1/p_\perp)=5\times10^{-5}\,\mathrm{GeV}^{-1}$$$$ \text{and}\hspace{1em} \sigma(r\phi)=\sigma(rz)=\left(4.2\oplus 4.0/(p\sin^{3/2}\theta)\right)\,\mu\mathrm{m}~.$$
The energy of the jets is smeared according to $\sigma(E)/E=0.3/\sqrt{E(\mathrm{GeV})}$~.
The position of the vertex is determined by the shape of the beam~\cite{teslaa} for the 
$x$ and $y$ coordinates and by the charged decay products of the Z for the $z$ coordinate.
The smearing of these coordinates is done accordingly.

\begin{figure}[t]
\begin{center}
    \epsfig{file=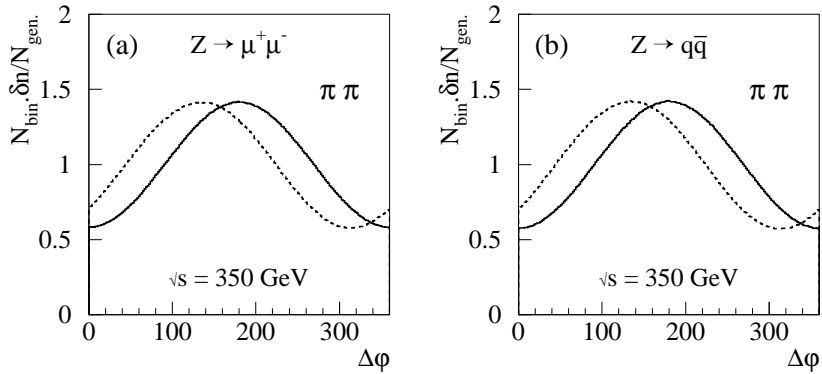,width=12cm}
    \caption{ The reconstructed distributions of $\Delta\varphi$ for the $\pi\pi$  channel at an energy
of 350 GeV, when all the experimental effects are taken into account: (a) for the decay of the Z into
$\mu^+\mu^-$, (b) for its decays into two jets. The convention of normalization and the values 
of $\psi$ are the same as in Fig.~\ref{fig:bsonly}\label{fig:pipi}.}

\end{center}
\end{figure}

The reconstructed distributions at an energy of 350\,GeV, with all the experimental effects 
included in the simulation, are shown in Fig.~\ref{fig:pipi} for the decays of the Z both into 
$\mu^+\mu^-$ and into $q\bar q$.  Their sensitivities to $\psi$ at the 
three considered energies are given in Table~\ref{tbl:pipismr}. 
It is clear from these values that a large part of the sensitivity is retained. 

Because of the key 
role of the vertex detector in the reconstruction, we have done again the simulations with 
the vertex detector errors multiplied by two. The reduction of the sensitivities that results 
from the increased  uncertainties is always smaller than 0.1.

\section{Conclusion}\longpage
We have studied the production of a light Higgs boson by the process of Higgsstrahlung and its
subsequent decay into $\tau^+\tau^-$, under the conditions of a linear collider.

We have described a method, which by the joint use of kinematics and vertexing allows the measurement
 of the transverse spin correlations of the two $\tau$'s.
This method is not impaired by beamstrahlung and can be applied for the main hadronic 
decay modes of the $\tau$ and  most of the visible decay modes of the Z. 

In the case of two $\tau^\pm\to\pi^\pm\bar{\nu}_\tau(\nu_\tau)$ decays, a complete simulation of the detector effects with the 
parameters of the TESLA project~\cite{teslab} has been performed.
A realistic study of 
the reconstruction of the $\pi^0$ is still to be done, nevertheless it appears that a reasonable goal for the
measurement of the phase $\psi$ that parametrizes a possible CP violation should be to 
use all the above mentioned final states and get a 
sensitivity better than 0.5, i.e., $\sigma_\psi<0.6\,\pi/\sqrt{N_{evt.}}$.


\begin{thebibliography}{10}

\bibitem{ssaa}
J.R.~Dell'Aquila and C.A.~Nelson,
Nucl. Phys. B320 (1989) 61. 
\bibitem{ssa}
C.A.~Nelson, Phys. Rev. D { 41} (1990) 2805.
\bibitem{ssb}M.~Kr\"amer, J.~K\"uhn, M.L. Stong, and P.M.~Zerwas,
Z. Phys. C { 64} (1994), 21.
\bibitem{ssc}
V.~Barger, K.~Cheung, A.~Djouadi, B.A. Kniehl, and  P.M.~Zerwas,
Phys. Rev. D { 49} (1994) 79.
\bibitem{ssd}
B.~Grz\c{a}dkowski and J.F.~Gunion, Phys. Lett. B { 350} (1995) 218.
\bibitem{ww}
 Z.~Was, and M.~Worek, Acta Phys. Polon. B33 (2002) 1875, hep-ph/0202007.
\bibitem{wa} G.R.~Bower, T.~Pierzcha{\l}a, Z.~W\c{a}s, and M.~Worek,
Phys. Lett. B { 543} (2002) 227.
\bibitem{wb} K.~Desch, Z.~W\c{a}s, M.~Worek,
Eur. Phys. J. C { 29} (2003) 491.
\bibitem{wc}K.~Desch, A.~Imhof, Z.~Was, and M.~Worek,
Phys. Lett. B { 579} (2004) 157.

\bibitem{omega}
{M.~Davier, L.~Duflot, F.~Le~Diberder and A.~Roug\'e,} { {Phys. Lett. B}}
  { 306} (1993) 411.
\bibitem{tsai}
Y.S.~Tsai, Phys. Rev. D 4 (1971) 2821.

\bibitem{kw}
H.~K\"uhn and F.~Wagner, Nucl. Phys. B236 (1984) 16.
\bibitem{analv}
S.~Jadach, J.H.~K\"uhn, and Z.~W\c{a}s,
Comput. Phys. Commun. { 64} (1991) 275.  

\bibitem{circe}
T.~Ohl, 
Comput. Phys. Commun. { 101} (1997) 269.  

\bibitem{tsh}
Y.-S. Tsai and A.C.~Hearn Phys. Rev. 140 (1965) B721.

\bibitem{polal}
{ ALEPH} Collaboration, A.~Heister et al., 
 Eur. Phys. J. C { 20} (2001) 401.

\bibitem{kuhn93}
{J.H.~K\"uhn,} { {Phys. Lett. B}} { 313} (1993) 458.

\bibitem{teslaa}
 TESLA Technical Design Report, Part II The Accelerator, 
R.~Brinkman, K.~Fl\"ottmann, J.~Ro{\ss}bach, P.~Schm\"user, N.~Walker, and
H.~Weise Edts.,
DESY~2001-011. 

\bibitem{teslab}
TESLA Technical Design Report, Part IV A Detector for TESLA,
T.~Behnke, S.~Bertolucci, R.D.~Heuer, and R.~Settles Edts., 
DESY~2001-011.

\end{thebibliography}
\end{document}